\documentclass[prd,email,showpacs,showkeys,preprintnumbers,amsmath,amssymb,nofootinbib]{revtex4}
\usepackage{amsmath}
\usepackage{latexsym}
\usepackage{graphicx}

\begin{document}

\title{Aspects of Magnetic Field Configurations in Planar
Nonlinear Electrodynamics}

\author{L. P. G. De Assis$^{1,3}$}
\email{lpgassis@ufrj.br}
\author{Patricio Gaete$^2$}
\email{patricio.gaete@usm.cl}
\author{Jos\'{e} A. Hela\"{y}el-Neto$^3$}
\email{helayel@cbpf.br}
\author{S. O. Vellozo$^{4,5}$}
\email{vellozo@cbpf.br}

\affiliation{${}^{1}$Departamento de F\'{\i}sica, Universidade
Federal Rural do Rio de Janeiro, BR 465-07, 23851-180,
Serop\'{e}dica, RJ, Brasil\\
${}^{2}$Departmento de F\'{\i}sica and Centro Cient\'{\i}fico-Tecnol\'{o}gico
de Valpara\'{\i}so, \\
Universidad T\'ecnica Federico Santa Mar\'{\i}a, Valpara\'{\i}so, Chile \\
${}^{3}$ Centro Brasileiro de Pesquisas F\'{\i}sicas,\\
Rua Xavier Sigaud, 150, Urca, 22290-180, Rio de Janeiro, Brasil\\
${}^{4}$ Centro Tecnologico do Ex\'{e}rcito-CTEx, Av. das Americas 28705,
230020-470, Rio de Janeiro, RJ, Brasil\\
${}^{5}$Instituto Militar de Engenharia (IME), Pra\c{c}a Gen. Tib\'{u}rcio s/n.
Praia Vermelha, Urca, Rio de Janeiro, RJ, Brasil \\
\today\\}



\begin{abstract}
\noindent In the framework of three-dimensional Born-Infeld
Electrodynamics, we pursue an investigation of the consequences
of the space-time dimensionality on the existence of
magnetostatic fields generated by electric charges at rest
in an inertial frame, which are present in its four-dimensional
version. Our analysis reveals interesting features of the model.
In fact, a magnetostatic field associated with an electric charge
at rest does not appear in this case. Interestingly, the addition
of the topological term (Chern-Simons) to Born-Infeld Electrodynamics
yields the appearance of the magnetostatic field. We also
contemplate the fields associated to the would-be-magnetic monopole
in three dimensions.
\end{abstract}

\maketitle

\pagestyle{myheadings}
\markright{{\it Aspects of Magnetic Field Configurations in Planar
Nonlinear Electrodynamics}}


\section{Introduction}
\renewcommand{\theequation}{1.\arabic{equation}}
\setcounter{equation}{0}

As well-known, nonlinear Born-Infeld Electrodynamics has become,
over the past years, a focus of intense activity after its connection
with brane physics has been elucidated. Specifically, the low-energy
dynamics of D-branes can be described by a nonlinear Born-Infeld-type
action \cite{Tseytlin,Gibbons}. We further recall that Born and Infeld
\cite{Born} suggested to modify Maxwell Electrodynamics to get rid of
infinities in the theory. Also, it is important to mention that, in addition
to the string interest, the Born-Infeld theory has also attracted attention
from different viewpoints. For example, in connection with noncommutative
field theories \cite{Gomis}, also in magnetic monopoles studies \cite{Kim},
and possible experimental determination of the parameter that measures the
nonlinearity of the theory \cite{Denisov}. More recently, it has been shown
that Born-Infeld Electrodynamics, due to its non-linearity, predicts the
existence of real and well-behaved magnetostatic
field associated with a electric charge at rest \cite{Helayel1,Helayel2}.
It is important to emphasize that this result is a non-linear effect restricted
to four space-time dimensions only, and it naturally raises the question about its
generalization to lower dimensions. Indeed, it is not quite evident that the
same phenomenon will take place in three space-time dimensions. The present
work specifically deals with this problem, where we examine the effect of
the space-time dimensionality on magnetostatic fields associated with electric
charges at rest.

On the other hand, it is also important to recall here that systems in
three-dimensional theories have been extensively discussed over the last few
years \cite{Deser,Dunne,Khare}. This is primarily due to the possibility
of realizing fractional statistics, where the physical excitations obeying it are
called anyons, which continuously interpolate between bosons and fermions. In this
connection, three-dimensional Chern-Simons gauge theory sets out as a standard
description, so that Wilczek's charge-flux composite model of anyon can be implemented
\cite{Wilczek}. Also, three-dimensional theories are interesting due to its
connection with the high-temperature limit of four-dimensional theories
\cite{Appelquist,Templeton,Das}, as well as for their applications to Condensed
Matter Physics \cite{Stone}.

In this perspective, it is of interest to improve our understanding of
the physical consequences of Born-Infeld Electrodynamics. As
already mentioned, of special interest will be to further explore the impact of
the space-time dimensionality on magnetostatic fields associated with
electric charges at rest. As we shall see, in the case of three-dimensional
Born-Infeld Electrodynamics, unexpected features are found. In fact, a
magnetostatic field associated with a electric charge at rest not appears in
this case. Interestingly, the addition of a topological term (Chern-Simons)
to Born-Infeld Electrodynamics results in the occurrence of novel non-trivial
solutions for the magnetic field, which clearly shows the key role played by
the Chern-Simons term. It is worth mentioning that these solutions agree with
those of Ref. \cite{Gaete}. Furthermore, our analysis goes on to explore what
is the three-dimensional analogue of a magnetic monopole. In other words,
what kind of electric charge-like is responsible for electrostatic fields with
non-vanishing rotational. The general outline of our paper is now described.
This work is a sequel to Ref. \cite{Helayel1,Helayel2} where we provide a
analysis of three-dimensional Born-Infeld electrodynamics by studying the
respective equations of motion. In Section II, we are most concerned with
properties of electrostatic and magnetostatic fields.

\section{Three-dimensional Born-Infeld electrostatic}\label{s21}
\renewcommand{\theequation}{2.\arabic{equation}}
\setcounter{equation}{0}

We now examine the equations of motion for Born-Infeld electrodynamics in three
dimensions, that is,

\begin{equation}
\overrightarrow{\nabla}\cdot\overrightarrow{D}=q\delta^{2}(\overrightarrow
{x})\rightarrow\overrightarrow{D}=\frac{q}{\rho}\widehat{\rho}, \label{4}
\end{equation}

\begin{equation}
\overrightarrow{\nabla}H=\overrightarrow{0}, \label{5}
\end{equation}
where the corresponding constitutive relations are given by

\begin{equation}
\overrightarrow{D}=\frac{\overrightarrow{E}}{\sqrt{1-\frac{\overrightarrow
{E}^{2}-B^{2}}{b^{2}}}}, \label{6}
\end{equation}

\begin{equation}
H=\frac{B}{\sqrt{1-\frac{\overrightarrow{E}^{2}-B^{2}}{b^{2}}}}. \label{7}
\end{equation}

It is worthwhile mentioning at this point that unlike to our previous analysis
\cite{Helayel1,Helayel2}, equation (\ref{5}) requires that the magnetic
field be always constant, leading to drastic consequences for the magnetic
sector, as it can be seen below. Accordingly, the magnetic induction, $B$,
present in the constitutive relation (\ref{7}) can be written as

\begin{equation}
B=\frac{H}{\sqrt{1-\frac{H^{2}}{b^{2}}}}\sqrt{1-\left(  \frac{\overrightarrow
{E}}{b}\right)  ^{2}}.
\end{equation}

We admit radial and/or angular dependence in the region of space where the
nonlinearity dominates. However, further analysis shows that, far enough
from the electric charge, so that the linear regime can be established,
$\left\vert \overrightarrow{E}\right\vert \ll b$ and $B\approx\frac{H}
{\sqrt{1-\frac{H^{2}}{b^{2}}}}$.

Since $B$ must tend to zero when
$r\rightarrow\infty$, this implies that $H$ must always be zero, because it
is constant. We remark that the new feature of the present model is that
the magnetic field generated from an standstill electric charge is ruled
out in three dimensions. This situation is exactly the opposite to that
described for Born-Infeld Electrodynamics in four dimensions.

\subsection{Three-dimensional topologically massive Born-Infeld
electrostatic}\label{s3}

The result above may suffer a drastic change whenever we allow that a topological term is added to Born-Infeld Electrodynamics. Thus, for a standstill point-like electric charge of intensity $q$, at $\rho=0$, the equations of motion are given by

\begin{equation}
\overrightarrow{\nabla}\cdot\overrightarrow{D}+\kappa B=q\delta(\rho), \label{7A}
\end{equation}
and
\begin{equation}
\overrightarrow{\nabla}H+\kappa\overrightarrow{E}=\overrightarrow{0}.
\label{7B}
\end{equation}
These equations, together with the constitutive relations (\ref{6}) and (\ref{7}),
describe the static and planar Born-Infeld electrodynamics coupled with a
term Chern-Simons. Here, $k$ is the coefficient of the Chern-Simons Theory.
We also recall that it is an integer multiple of some basic unit in the quantum
Hall effect (QHE), while in the fractional quantum Hall effect (QHE) it is a
rational multiple.

Considering the purely radial dependence and only a single radial component for the
fields $\overrightarrow{E}$ and $\overrightarrow{D}$, these two equations can be
written as

\begin{equation}
\frac{1}{\rho}\frac{d}{d\rho}\left(  \rho D\right)  +\kappa B=q\delta(\rho), \label{8}
\end{equation}

\begin{equation}
\frac{d}{d\rho}\left(  H\right)  +\kappa E=0. \label{9}
\end{equation}
Moreover, the constitutive relations (\ref{6}) and (\ref{7}), for the $E$ and $B$
fields read

\begin{equation}
B=\frac{H}{\sqrt{1+\frac{D^{2}}{b^{2}}-\frac{H^{2}}{b^{2}}}}, \label{9.1}
\end{equation}

\begin{equation}
E=\frac{D}{\sqrt{1+\frac{D^{2}}{b^{2}}-\frac{H^{2}}{b^{2}}}}. \label{9.2}
\end{equation}
With the help of (\ref{9.1}) and (\ref{9.2}), equations (\ref{8}) and (\ref{9}) can be rewritten as

\begin{equation}
\frac{1}{\rho}\frac{d}{d\rho}\left(  \rho D\right)  +\frac{\kappa H}
{\sqrt{1+\frac{D^{2}}{b^{2}}-\frac{H^{2}}{b^{2}}}}=q\delta(\rho), \label{10}
\end{equation}

\begin{equation}
\frac{d}{d\rho}\left(  H\right)  +\frac{\kappa D}{\sqrt{1+\frac{D^{2}}{b^{2}
}-\frac{H^{2}}{b^{2}}}}=0. \label{11}
\end{equation}

We observe that this system has no analytical solution all over the space. In the
region far from the electric charge, the fields are weak and the solution is
already known

\begin{equation}
D=E=aK_{1}(\kappa\rho), \label{11AA}
\end{equation}

\begin{equation}
H=B=aK_{0}(\kappa\rho). \label{11AB}
\end{equation}

Close to the electric charge, the fields are strong and the
nonlinearity is present. The fields $E$ and $D$, for example, are no longer
proportional.

It is important to realize that equation (\ref{11}) states that the derivative
of the scalar magnetic field, $H$, is always negative, if $D$ and $k$ are positive.
Therefore is a decreasing function. At this point, we make the assumption that the
field $D$ is singular at $\rho=0$. Thus, the term inside the root ($1+\frac{D^{2}}{b^{2}}-\frac{H^{2}}{b^{2}}$) will be dominated by $\frac{D^{2}}{b^{2}}$
and near the origin the derivative of the field $H$ will reduces to
\begin{equation}
\frac{d}{d\rho}\left(  H\right)  \approx-\frac{\kappa D}{\sqrt{\frac{D^{2}
}{b^{2}}}}=-\kappa b.
\end{equation}

By assuming that $H(0)=Ho$ (value to be determined), the field $H(\rho)\approx
Ho-kb\rho$ when $\rho\approx0$. Putting this result into equation
(\ref{10}) (near the origin), we arrive at the following differential equation
for the field $D$
\begin{equation}
\frac{1}{\rho}\frac{d}{d\rho}\left(  \rho D\right)  =-\frac{\kappa b\left(
H_{o}-\kappa b\rho\right)  }{D}.
\end{equation}
This allows us to find the following solution for $\rho\approx0$
\begin{equation}
D(\rho)\approx\frac{C}{\rho}.
\end{equation}
In other terms, this confirms the $\frac{1}{\rho}$-singularity for the $D$-field.
The constant $C$ is related to the electric charge if one takes the $2\pi\rho D(\rho)=q$.
In order to estimate the behavior of the $B$-field in this region, we examine equation (\ref{9.1}), that is,
\begin{equation}
B=\frac{H}{\sqrt{1+\frac{D^{2}}{b^{2}}-\frac{H^{2}}{b^{2}}}}\approx\frac
{H_{o}-\kappa b\rho}{\sqrt{\frac{D^{2}}{b^{2}}}}=\frac{b\left(  H_{o}-\kappa
b\rho\right)  \rho}{C}\rightarrow0.
\end{equation}

Here, we see the strong nonlinear effect of the fields near the electric
charge. While $H$ is finite and nonzero, the field $B$ vanishes due to the
singular behavior of the field $D$.

Equation (\ref{9.2}) we can evaluate the electric field
$E$ near the origin. The field $E$ goes to $b$, as the square
root of the denominator is dominated by the field $D$. As a result, the electric
field is not only finite but it also assumes the maximum Born-Infeld field intensity,
$b$, totally different from the $D$ singular field behavior.

\begin{equation}
E=\frac{D}{\sqrt{1+\frac{D^{2}}{b^{2}}-\frac{H^{2}}{b^{2}}}}\approx\frac
{D}{\sqrt{\frac{D^{2}}{b^{2}}}}=b
\end{equation}

Finally, we can determine the value of the magnetic field $H$ at the origin
using the constitutive relation (\ref{7}):
\begin{equation}
H=\frac{B}{\sqrt{1-\frac{\overrightarrow{E}^{2}-B^{2}}{b^{2}}}}\rightarrow
H=\frac{B}{\sqrt{\frac{B^{2}}{b^{2}}}}\rightarrow b
\end{equation}

A summary of the behavior of fields will provide an overview of their spatial
distribution and guide us to the correct boundary values on numerical solution
of the problem. Thus, we have the following picture:

For $\rho\rightarrow\infty$, the behavior is linear and well established

\begin{center}
\begin{equation}
D=E=aK_{1}(\kappa\rho),
\end{equation}

\begin{equation}
H=B=aK_{0}(\kappa\rho).
\end{equation}

\end{center}

For $\rho\rightarrow0$, the fields radically differ, showing the effects of the
non-linearity

\begin{equation}
E\rightarrow b,
\end{equation}

\begin{equation}
D\rightarrow\infty,
\end{equation}

\begin{equation}
H\rightarrow b,
\end{equation}

\begin{equation}
B\rightarrow0.
\end{equation}

\subsection{Setting bounds}\label{s22}
\renewcommand{\theequation}{2.2.\arabic{equation}}
\setcounter{equation}{0}

We have the following challenge: the fields $E$ and $H$ at $\rho=0$ is unknown
because we ignore the maximum field strength $b$. We must use a different strategy.
Far away from the electric charge, where the linear regime dominates, the fields are well described by analytical functions (\ref{11AA}) and (\ref{11AB}). The Bessel functions describe the shape of the fields. The constant $a$ describes the field amplitude and
it can be estimated with good accuracy. The intention is to use that field as a
boundary condition. For this purpose, the phenomenon of superconductivity can be
very useful.  It has a quasi-planar structure and can be described by a three-dimensional model. The analogy is present also in its linear equation for the magnetic field.
Many physical properties of high-temperature superconductors are two-dimensional
phenomena derived from their square planar crystal structure. Interest in planar
field theories is mainly motivated by the possibility of understand the critical
phenomena and extend them to four-dimensions. How we know, the hallmark of
superconductivity is the expulsion of the $B$ field from the inner region of the superconductor, called Meissner Effect. The London penetration depth, $L$, is well experimentally measured for several super-conducting materials. Its value ranges
from $10$ to $100$ nm. Therefore, these experimental data are of great importance
for an indirect estimate of limits or ranges for the BI maximum field strength $b$,
if it exist. Our equations, in the linear regime, are exactly the same if the
CS-coefficient be associated with this penetration length, $k=1/L$. Our task is to
solve numerically the system of equations (\ref{8}) and (\ref{9}), look for fields
spatial distribution until nonlinear region, satisfying the previously established
limits at the $\rho=0$ and finally, take the $b$ directly from the curve.

To calculate the amplitude, $a$, of the electric field, we take a typical value
involving the charge and the characteristic length as $\frac{q}{4\pi\epsilon_{0}{} L}$,
where $\epsilon_{0}$ is the permittivity of free space and is numerically equal to $8.854\times 10^{-12} \:\frac{C^{2}}{Nm^{2}}$. The charge $q$ has $C/m$ units and
will be set as $e/L$, where $e$ is the elementary charge. The maximum experimental characteristic length, $L$, which is around $100\:nm$, will set up the lower limit
for maximum field $b$ of Born-Infeld Theory. With these data set the estimated value
for $a$ is equal to $1.4 \times 10 ^{5}$. The result can be seen in the Figure $1$.

\begin{figure}
[h]
\begin{center}
\includegraphics[angle=0,width=0.80\textwidth]
{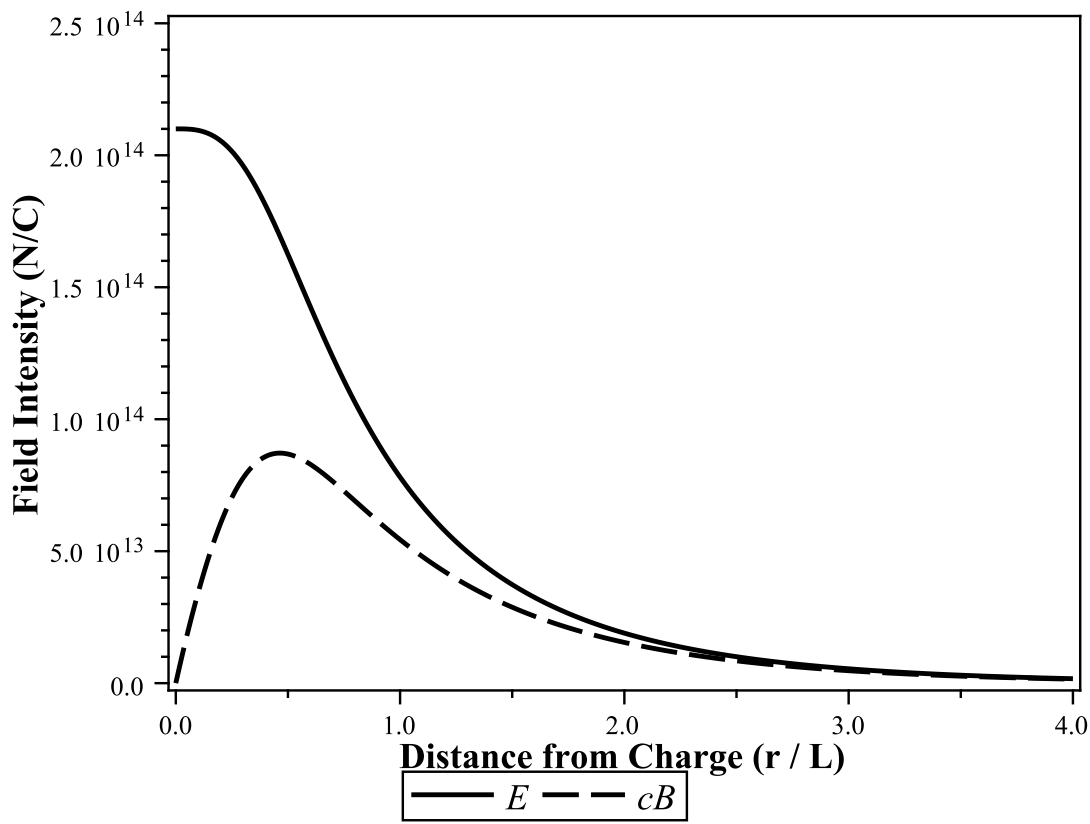}%
\end{center}
\caption{\label{fig1} \it The fundamental fields near the electric charge.}
\end{figure}

These curves confirm the prior analysis predictions for the fields far away
and near of the electric charge. An alternative setting was also performed
in \cite{Manoel1,Manoel2}.

\section{Bianchi identity violation for three-dimensional topologically
massive Born-Infeld Electrodynamics}\label{s4}
\renewcommand{\theequation}{3.\arabic{equation}}
\setcounter{equation}{0}

The possibility of Bianchi identity violation is considered. This section explores
that consequences in the context of Born-Infeld with topological Chern-Simons term
added. In the static and planar version, it acquires the following form:

\begin{equation}
\overrightarrow{\nabla}\times\overrightarrow{E}=\overrightarrow{\nabla}%
\times(-\overrightarrow{\nabla}\phi)=\sigma\delta^{2}(\overrightarrow{x}). \label{20}
\end{equation}

This equation requires that the scalar potential $\phi$ be non regular, in analogy
to the four dimensional case where the potential vector, $\overrightarrow{A}$, exhibits
a string of singularities, allowing the introduction of magnetic
charges. Writing equation (\ref{20}) in Cartesian coordinates,

\begin{equation}
\varepsilon^{ij}\partial_{i}\partial_{j}\phi=-\sigma\delta^{2}(\overrightarrow
{x})\rightarrow\left(  \partial_{x}\partial_{y}-\partial_{y}\partial
_{x}\right)  \phi=-\sigma\delta(x)\delta\left(  y\right),
\end{equation}

so that the solution is well-known and is written as:

\begin{equation}
\phi=-\frac{\sigma}{2\pi}tg^{-1}\left(  \frac{y}{x}\right)  .
\end{equation}

Changing to polar coordinates the solution acquires an immediate interpretation:

\begin{equation}
\phi=-\frac{\sigma}{2\pi}\theta,
\end{equation}

As a consequence, we have a tangential electric field that arises from an
scalar current:

\begin{equation}
\overrightarrow{E}=\frac{\sigma}{\rho}\widehat{\theta}. \label{1}
\end{equation}

Taking the BICS equation for $H$ and $\overrightarrow{E}$:

\begin{equation}
\overrightarrow{\nabla}H+\kappa\overrightarrow{E}=\overrightarrow{0}%
\end{equation}

\begin{equation}
\overrightarrow{\nabla}H=-\kappa\frac{\sigma}{\rho}\widehat{\theta}%
\end{equation}

The solution is immediate:

\begin{equation}
H=-\kappa\sigma\theta+C \label{2}%
\end{equation}

Inverting the constitutive relation of BI we have:

\begin{equation}
H=\frac{B}{\sqrt{1-\frac{\overrightarrow{E}^{2}-B^{2}}{b^{2}}}}\rightarrow
B=\frac{H}{\sqrt{1-\frac{H^{2}}{b^{2}}}}\sqrt{1-\left(  \frac{\overrightarrow
{E}}{b}\right)  ^{2}} \label{2.1}%
\end{equation}

Assuming $C=0$, and substituting (\ref{1}) and (\ref{2}) in (\ref{2.1}),
we arrive at:%

\begin{equation}
B=-\frac{\kappa\sigma\theta}{\sqrt{1-\frac{\left(  \kappa\sigma\theta\right)
^{2}}{b^{2}}}}\sqrt{1-\left(  \frac{\sigma}{\rho b}\right)  ^{2}} \label{3}%
\end{equation}

This equation for the field $B$ is dependent on $\rho$ and $\theta$. Far
away the electric charge, the field $B$ behaves like:%

\begin{equation}
B\approx-\frac{\kappa\sigma\theta}{\sqrt{1-\frac{\left(\kappa\sigma
\theta\right)^{2}}{b^{2}}}}%
\end{equation}

Assuming $\kappa\sigma\theta\ll b$, we restore the limit
$B\approx H$. However, the argument of the square root
$\sqrt{1-\left(  \frac{\sigma}{\rho b}\right) ^{2}}$, when close to the
electric charge, can become negative leaving the field $B$ complex.
In addition, it exhibits the same discontinuity of the field $H$ when the
angle $\theta$ is close to $2\pi$.

\section{Final Remarks}

Our main effort in the present paper has been to understand how non-linear
effects of planar Electrodynamics in Born-Infeld formulation may induce magnetic
effects out of simple electrostatic systems.

Lower dimensionality actually reveals special properties if compared with the
same problem in its $(3+1)$-D formulation \cite{Helayel1,Helayel2,Botta}. However, we
would like to formulate as much as possible how to understand the results of
this work in connection with the dimensional reduction of $(3+1)$-D Electrodynamics
to $(2+1)$-D.

Both planar situations studied here,
${\nabla} \cdot {\bf E} = q \delta ^{\left( 2 \right)} \left( {\bf x} \right)$
and ${\nabla}  \times {\bf E} = \sigma \delta ^{\left( 2 \right)} \left( {\bf x} \right)$, can be mapped into the (spatial) $3$-dimensional problem of calculating the magnetostatic field on an infinite wire placed along the $z$-axis with a steady current density,
${\bf j} = I_0 \delta \left( x \right)\delta \left( y \right)\hat z$. If
${\bf {\cal E}} = \left( {{\bf E};E \equiv E_z } \right)$ and
${\bf {\cal B}} = \left( {{\bf B};B \equiv B_z } \right)$ stand for the electric and magnetic fields in $3$ spatial dimensions (where ${\bf E}$ and ${\bf B}$ denote vectors in $2$ space dimensions $(xy-plane)$), planar Maxwell Electrodynamics rules out $E$ and $\bf B$
and we have Maxwell equations for ${\bf E}$ and $B$ in $(2+1)$-D:
\begin{equation}
{\nabla}  \cdot {\bf E} = \rho , \label{FR5}
\end{equation}
\begin{equation}
{\nabla}  \times {\bf E} =  - \partial _t B, \label{FR10}
\end{equation}
\begin{equation}
{\Tilde \nabla} B = \frac{1}{{c^2 }}\partial _t {\bf E} + \mu _0 {\bf j}, \label{FR15}
\end{equation}
where the $\nabla$-operator, $\bf E$ and $\bf j$ are planar vectors and $B$ behaves as a pseudo-scalar in ${\cal R}^2$. Planar Maxwell Electrodynamics singles out the fields above and discard the following equations for $E$ and ${\bf B}$:
\begin{equation}
\tilde \nabla E - \partial _z \tilde {\bf E} =  - \partial _t {\bf B}, \label{FR20}
\end{equation}
and
\begin{equation}
\nabla  \times {\bf B} = \frac{1}{{c^2 }}\partial _t  E + \mu _0 j, \label{FR25}
\end{equation}
where $\tilde \nabla  = \left( {\partial _y ; - \partial _x } \right)$,
$\tilde {\bf E} = \left( {E_y ; - E_x } \right)$, $\tilde {\bf B} = \left( {B_y ; - B_x } \right)$ and $E$ and $j$ are scalars in ${\cal R}^2$.

Then, a point-like charge at the origin in the planar case is not the dimensional reduction of the corresponding situation in ${\cal R}^3$. It rather corresponds to the case of an infinite wire with a steady current in ${\cal R}^3$:
${\bf j} = \left( {0;0;j = I_0 \delta \left( x \right)\delta \left( y \right)} \right)$.
Thus, the planar equations ${\nabla} \cdot {\bf E} = q\delta ^{\left( 2 \right)} \left( {\bf x} \right)$
and ${\nabla}  \times {\bf E} = \sigma \delta ^{\left( 2 \right)} \left( {\bf x} \right)$
can be both re-expressed in terms of the equation
$\nabla  \times {\bf B} = \mu _0 j = I_0 \delta \left( x \right)\delta \left( y \right)$,
($\partial_t E=0$) which can be written in $2$ spatial dimensions as
$\nabla  \cdot \tilde {\bf B} = \mu _0 I_0 \delta ^2 \left( {\bf x} \right)$, whose solution (in polar coordinates) is $\tilde {\bf B} = \frac{{\mu _0 I_0 }}{{2\pi \rho}}\hat r$, so that
${\bf B} = \frac{{\mu _0 I_0 }}{{2\pi \rho}}\hat \theta$. We stress that this ${\bf B}$-field is ruled out by planar Maxwell Electrodynamics. It is the $2$-dimensional sector of the $3$D magnetic field.

So, in the case $\nabla  \cdot {\bf E} = \frac{q}{{\varepsilon _0 }}\delta ^{\left( 2 \right)} \left( {\bf x} \right)$, we see that ${\bf E}$ replaces $\tilde {\bf B}$ in $\nabla  \cdot \tilde {\bf B} = \mu _0 I_0 \delta ^2 \left( {\bf x} \right)$,
whereas in the case $\nabla  \times {\bf E} = \nabla  \cdot \tilde {\bf E} = \sigma \delta ^{\left( 2 \right)} \left( {\bf x} \right)$, it is ${\tilde {\bf E}}$ that plays the r\^{o}le of ${\tilde {\bf B}}$. Then, we may say that the electrostatic fields we have in planar Electrodynamics correspond to the ${\bf B}$-field ruled out by the low dimensionality.

We can now confirm why planar Electrostatics cannot support a $B$-field even in the non-linear case, as we have worked out in Section II-A. If the electric charge at rest in $2$D corresponds to the infinite wire with current in $3$D, the Born-Infeld equation for H reads $\tilde \nabla H - \partial _z \tilde {\bf H} = \mu _0 {\bf j}$, $\partial _t {\bf E} =0$. But ${\bf j}=0$, since only $j_z  = I_0 \delta \left( x \right)\delta \left( y \right)$ is non-trivial. Also, $\partial_z {\bf \tilde H}=0$, due to the translational symmetry along the $z$-axis. Then,
$\tilde \nabla H = 0$ and so $H=0$, since it must be constant and there are no sources at infinity.

Since in $3$D $ {\bf {\cal E}} \cdot  {\bf {\cal B}} = {\bf E} \cdot {\bf B} + EB$, and planar Electrodynamics rules out ${\bf B}$ and $E$, the vanishing of $H$ is equivalent to the vanishing of $B$. This is how we justify the absence of a $B$-field in non-linear Electrodynamics for a charge at rest, contrary to what happens in $3$D.

We shall exploit further this interplay between $(2+1)$-D and $(3+1)$-D Born-Infeld Electrodynamics focusing our attention on the problem of magnetic monopoles and vortices.

\section{Acknowledgments}
One of us (PG) wants to thank the Field Theory Group of the CBPF for hospitality
and PCI/MCT for support. This work was supported in part by Fondecyt (Chile)
grant 1080260 (PG). S.O.V. wish to thanks to CBPF and IME for Academic support
and CTEx by the financial help. L.P.G.A is grateful to FAPERJ-Rio de Janeiro for his post-doctoral fellowship. J.A.H.-N. expresses his gratitude to CNPq for financial help.

\end{document}